\documentclass[ aps,prd,preprint, tightenlines, nofootinbib,showpacs,fixfloat,superscriptaddress]{revtex4-1}
\usepackage{epsfig}
\usepackage{graphicx}
\usepackage{dcolumn}
\usepackage{bm}
\usepackage{overpic}
\usepackage{subfigure}
\usepackage{float}

\begin{document}


\title{Improved Energy Reconstruction in NOvA with Regression Convolutional Neural Networks}

\author{Pierre Baldi, Jianming Bian, Lars Hertel, Lingge Li}
\affiliation{%
 University of California, Irvine\\
 Irvine, California 92697, USA
}%



\begin{abstract}

In neutrino experiments, neutrino energy reconstruction is crucial because neutrino oscillations and differential cross-sections are functions of neutrino energy. It is also challenging due to the complexity in the detector response and kinematics of final state particles. We propose a regression Convolutional Neural Network (CNN) based method to reconstruct electron neutrino energy and electron energy in the NOvA neutrino experiment. We demonstrate that with raw detector pixel inputs, a regression CNN can reconstruct event energy even with complicated final states involving lepton and hadrons. Compared with kinematics-based energy reconstruction, this method achieves a significantly better energy resolution. The reconstructed to true energy ratio shows comparable or less dependence on true energy, hadronic energy fractions, and interaction modes. The regression CNN also shows smaller systematic uncertainties from the simulation of neutrino interactions. The proposed energy estimator provides improvements of $16\%$ and $12\%$ in RMS for $\nu_e$ CC and electron, respectively. This method can also be extended to solve other regression problems in HEP, taking over kinematics-based reconstruction tasks.


\end{abstract}

\maketitle


\section{Introduction}

Energy reconstruction plays a key role in High Energy Physics (HEP), as it converts detector unit readout into kinematics of interactions. In a HEP analysis, event and particle types are usually identified first and then reconstructed energies are assigned to individual final state particles and the overall event via the energy reconstruction process. Based on these energies, physical phenomena can be studied as functions of overall energy and internal kinematics of an event. In neutrino physics, energy reconstruction is essential and challenging for the neutrino oscillation studies. Traditionally, particle energies are reconstructed by adding up or fitting to the hits on detector readout units, and event energy is reconstructed as a function of particle energies in the event. In this paper, a deep-learning based method for energy reconstruction of neutrino oscillations will be discussed. The method directly uses detector hits as inputs without intermediate steps.

Neutrino oscillations are so far the only experimental observation beyond the standard model since its development about 30 years ago. Neutrinos are very elusive since they only interact via the weak nuclear force. They have three active flavor states $\nu_e$, $\nu_\mu$ and $\nu_\tau$. Each is a different superposition of three mass states $\nu_1$, $\nu_2$ and $\nu_3$. Neutrinos can oscillate between flavor states. The relationship between flavor and mass states, and the oscillation between flavors are commonly described by the Pontecorvo$-$Maki$-$Nakagawa$-$Sakata (PMNS) matrix \cite{ref:pmns}.

Two fundamental questions of interest are remaining to be determined by studying neutrino oscillations. First, what is the CP phase $\delta$? The CP phase $\delta$ relates to the difference in oscillation behavior between neutrinos and anti-neutrinos. CP violation in the lepton sector holds implications for matter-antimatter asymmetry in the Universe through leptogenesis. Second, what is the mass ordering ($m_3>m_{1,2}$ or $m_{1,2}>m_3$) between neutrinos? The mass hierarchy provides key information for future searches of the neutrino-less double beta decay. Observing the neutrino-less double beta decay would imply that the neutrino is a Majorana particle meaning it is its own anti-particle. The mass hierarchy will also constrain the so far undetermined absolute neutrino masses.

Aiming to solve these two questions, current and future neutrino oscillation experiments focus on electron neutrino appearance ($\nu_\mu\to\nu_e$). Neutrino oscillation can be measured by sending a beam of neutrinos of one flavor through a detector. Oscillated neutrinos arriving in the detector will be of a different flavor than the one generated from the beam. Observed neutrino interactions need to be tagged by their flavors. Importantly, the energy of the incoming neutrino needs to be well reconstructed as the $\nu_\mu\to\nu_e$ oscillation probability changes as a function of neutrino energy.

One of the challenges is thus a good estimation of electron neutrino energy. The accuracy of neutrino energy reconstruction influences how precisely neutrino oscillation parameters can be estimated.
An electron neutrino can only be identified in charged current (CC) interactions where the electron neutrino converts into an electron. The $\nu_e-CC$ events are characterized by an electron along with other potential activity produced by hadrons. Traditionally, the reconstructed neutrino energy is calculated as a function of electron and hadron visible energy deposits. However, the estimation of the energy with the kinematics based method is complicated by missing energy in dead material, non-linear detector energy responses, invisible energy and identities (mass) of hadrons, and overlaps between electron and hadron showers.

To address these issues, the neutrino energy can be predicted directly from images of interactions. These images provide additional information on interaction details such as trajectories and energy deposit patterns of electrons and hadrons. Deep learning and deep convolutional neural network methods are a natural choice for processing data produced by complex detectors in high energy physic~\cite{Shimmin:2017mfk,Sadowski:2017ilo,Baldi:2016fql,Baldi:2014kfa} and have demonstrated success in {\bf classification} problems in collider and neutrino experiments. NOvA has pioneered this deep learning technique in flavor tagging problems and has used it to produce oscillation physics results ~\cite{Aurisano:2016jvx,Adamson:2017gxd}. CNN based event identification and reconstruction have also been investigated in other neutrino experiments~\cite{Racah:2016gnm,Acciarri:2016ryt,Renner:2016trj,Perdue:2018ihs,Delaquis:2018zqi}. In this work, we propose to develop a {\bf regression} CNN based method to precisely reconstruct electron neutrino energy and electron energy in the NOvA neutrino experiment. Neutrino interactions at NOvA typically involve multiple final state particles with complicated kinematics. The convolutional filters in CNNs can extract a richer set of features of these events than the sum of energy deposits. For example, two neutrinos could deposit almost exactly the same amount of energy in the detector and convolutional neural network features learned from event topologies could be used to make a more refined prediction on the true neutrino energy than the  kinematics based method.

Reconstruction of individual final state particle energy in an interaction is a basic task of energy reconstruction. Specifically, these particle energies are used to study the kinematics, such as the momentum and energy transfer, in neutrino interactions. To demonstrate that the regression CNN can also reconstruct single particle energy, a similar method to the electron neutrino energy estimator is used to estimate electron energy. This estimator can be used for cross-section measurements and shower reconstruction study.

\section{The NOvA Experiment}\label{sec:novaexp}
NOvA uses an intense neutrino beam and sends it through sensitive, fine-grained detectors for long periods of time. The NuMI (Neutrinos at the Main Injector) muon neutrino beam is produced at Fermilab, Illinois. Aimed at 3.3 degrees downward, the beam travels 810 km through the earth to the 14 kilotons far detector (FD) in Ash River, Minnesota \cite{Adamson:2015dkw}. The far detector measures electron neutrinos oscillated from muon neutrinos in the beam. The NOvA experiment has the longest beam-detector distance in the world which maximizes the matter effect and allows a measurement of the neutrino mass ordering. Additionally, a 330 ton functionally identical near detector at Fermilab measures unoscillated beam neutrinos and estimates backgrounds and signals at the far detector. Both detectors are located 14 milli-radians off the centerline of the neutrino beam. This allows the detectors to capture a narrow energy spectrum of neutrinos at approximately 2 GeV. This is the energy at which the oscillation probability from a muon neutrino to an electron neutrino is expected to be at its peak.

The NOvA detectors are constructed in layers of alternating vertical and horizontal PVC cells; activities in the cells are recorded in a top view and a side view. There are 344,064 cells in the far detector and 18,000 cells in the near detector. In the far detector, each cell is 3.9 cm wide, 6.0 cm deep and 15.6 m long. The cells are made of highly reflective plastic filled with liquid scintillator. The scintillation light produced by neutrino interactions in the detectors is collected by a wavelength shifting fiber connected to an avalanche photodiode installed on one end of each cell. The readouts from these photodiodes are converted to calorimetric energy for physics analyses.


\section{Methods}\label{sec:method}

\subsection{Simulated Data Sample}\label{sec:sample}
\subsubsection{Simulation}

The standard NOvA simulation is used to generate training and validation samples for the regression CNN. The simulation of NuMI neutrino beam is described in Ref.~\cite{NOvA:2018gge}. The beam is simulated by GEANT4~\cite{Agostinelli:2002hh}  and corrected according to external thin-target hadroproduction data with the PPFX tool~\cite{Aliaga:2016oaz}.  The flux shape of the NuMI neutrino beam at NOvA is referred to as the regular flux in this paper.  Since NOvA is an off-axis experiment, the neutrino spectrum at the NOvA far detector from the NuMI flux peaks at about $2$ GeV, close to the $\nu_\mu\to\nu_e$ oscillation maximum. Since there are few low energy neutrino ($<$ 1 GeV) events in the NOvA FD Monte Carlo sample with the regular NuMI flux, the regression CNN $\nu_e$ energy trained with it has a significant true energy dependence (see Section~\ref{sec:result}). To minimize the dependence of estimated neutrino energy on true neutrino energy in the $\nu_e$ energy training, a flat neutrino flux shape is used to generate the far detector $\nu_e$ CC Monte Carlo sample to train the regression CNN for $\nu_e$ energy reconstruction. In the case of electron shower energy estimation, electrons from the regular flux $\nu_e$ CC far detector Monte Carlo sample are used in the electron energy regression CNN training and its validation.


At NOvA, interactions of neutrinos on nuclei are simulated by GENIE~\cite{Andreopoulos:2009rq}, and detector responses are then simulated by GEANT4. The customized NOvA detector simulation chain is described in~\cite{Aurisano:2015oxj}.

To study the $\nu_e$ interactions in the far detector Monte Carlo, we generate $\nu_e$ interactions with energies taken from the $\nu_\mu$ flux distribution. After that, no neutrino oscillations are applied to the training samples. This is equivalent to assuming all muon neutrinos oscillate to electron neutrinos in the far detector.  To study realistic energy reconstruction performances from different energy estimators, the real $\nu_e$ appearance signal in the far detector can be obtained by applying realistic oscillation weights to this sample.

The simulation produces image pairs of the entire detector. As explained in Section~\ref{sec:novaexp} the two images correspond to cells in the top view (X-view) planes and side view (Y-view) planes of the detector.  The images have a size 896$\times$384 (horizontal $\times$ vertical), where each pixel corresponds to the energy deposited in the corresponding detector cell. The horizontal coordinate (0-895) of a pixel represents the plane index of the detector cell and the vertical coordinate (0-383) represents the cell index in that plane. Since X-view planes and Y-view planes are assembled alternatively, all pixels with odd (even) plane indices are set to zero for X-View (Y-View).  The neutrino flavor and the type of interaction are tagged by the true neutrino interaction information in GENIE.

\subsubsection{Reconstruction}
The overall reconstruction process at NOvA is described in \cite{ref:reco0}. First, different neutrino interactions captured in the same pair of detector views are separated~\cite{ref:reco1}. Cell hits are clustered by space and time. This separates neutrino interactions caused by beam neutrinos from cosmic ray neutrinos in a time window. The procedure collects cell hits from a single neutrino interaction (slice). The slices then serve as the foundation for all later reconstruction stages. We will refer to one slice as neutrino interaction from here on.

For each neutrino interaction, the vertex is then identified. The vertex is where the neutrino interacts with the detector material. All particles created in the interaction originate at the vertex. In order to reconstruct the vertex position, a modified Hough transform is used to fit straight-lines to cell hits. Then the lines are tuned in an iterative procedure until they converge to the image's reconstructed vertex~\cite{ref:reco2, ref:reco3, ref:reco4, ref:fuzzyk, Niner:2015aya}. The cell closest to the reconstructed interaction vertex in each view is chosen as the reference cell in our pixel maps. We will refer to the reference cell as the reconstructed vertex from here on.

Since the size of the neutrino interaction is much smaller than the entire detector, the image can be cropped. An image can be cropped by considering the number of pixels that are occupied in all four directions from the reconstructed vertex. To determine a window size the distribution of electron neutrino interactions is inspected.
The cropped image contains 30 pixels to the left and 120 pixels to the right of the vertex. In the vertical-direction, 70 pixels above and below the vertex are included. This produces images of $151 \times 141$ pixels in each view. On average, $99.5\%$ of the hits are contained in a cropped image.

\begin{figure}[t]
  \centering
    \begin{minipage}[b]{0.8\linewidth}
    \includegraphics[width=\linewidth]{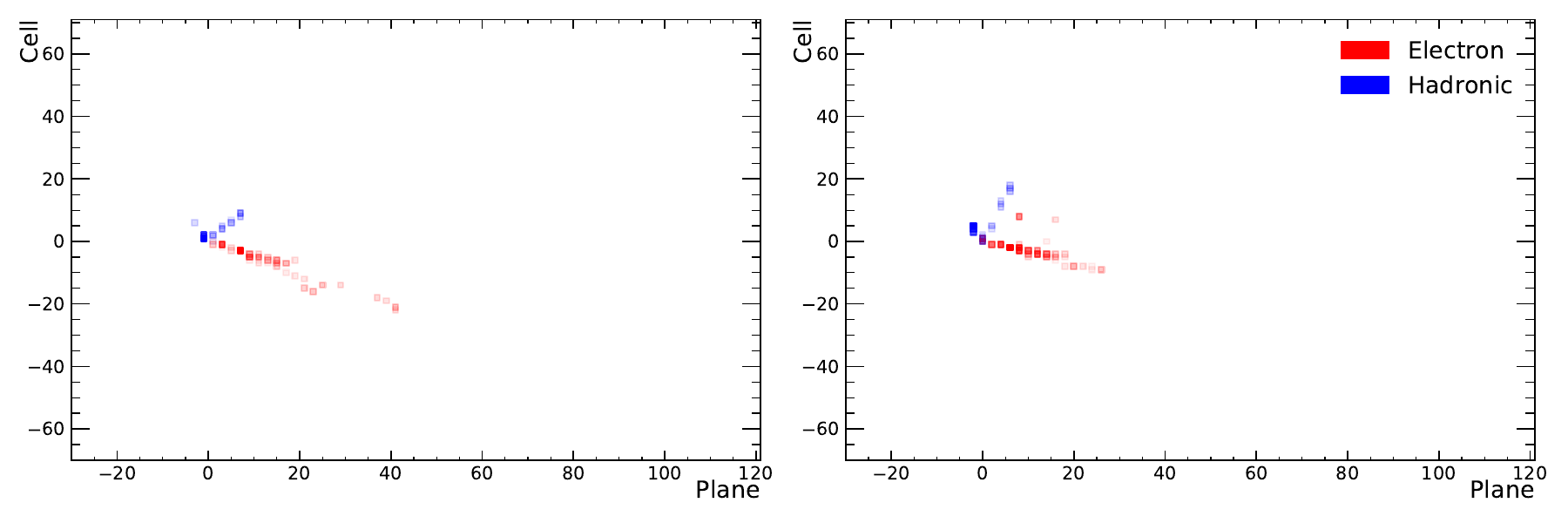}
  \end{minipage}
  \caption{Electron neutrino image pair example.}
    \label{fig:nue_image}
\end{figure}

  \begin{figure}[t]
    \centering
    \includegraphics[width=8cm]{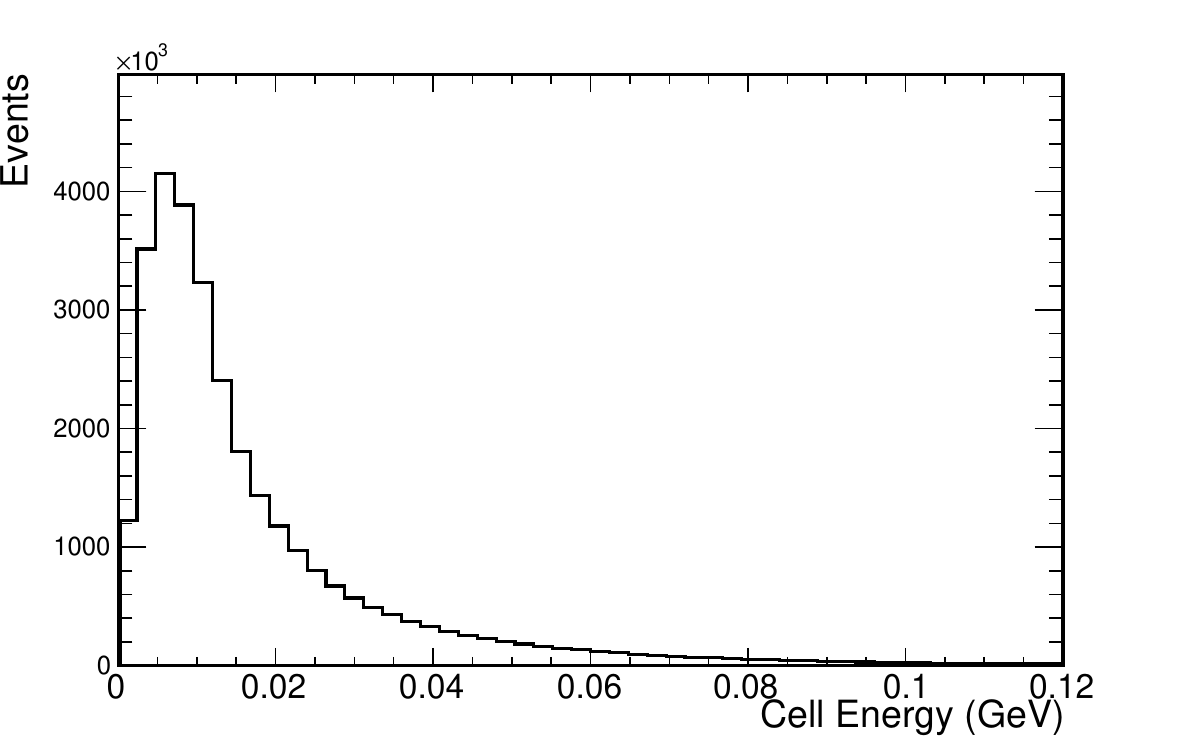}
     \includegraphics[width=8cm]{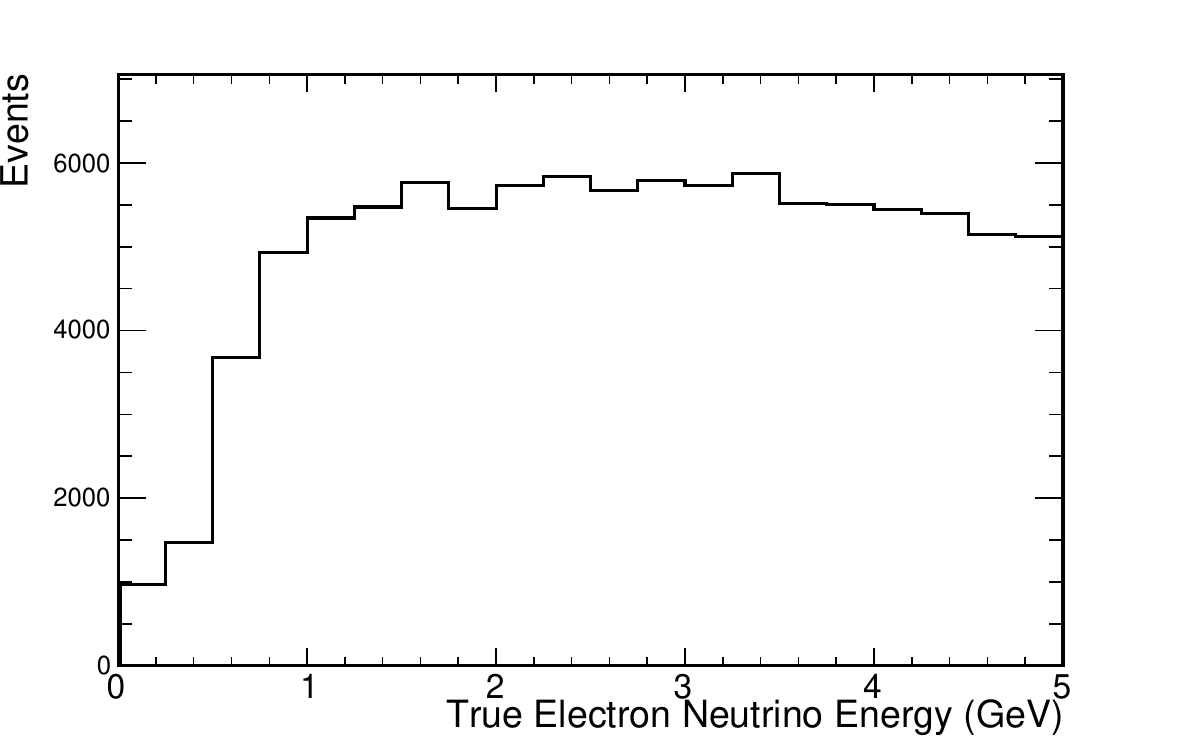}
    \caption{Distributions of input cell energies (left) and true electron neutrino energies (right) from a subset of the flat flux training sample.}
    \label{fig:nue_trainE_flat}    
    \includegraphics[width=8cm]{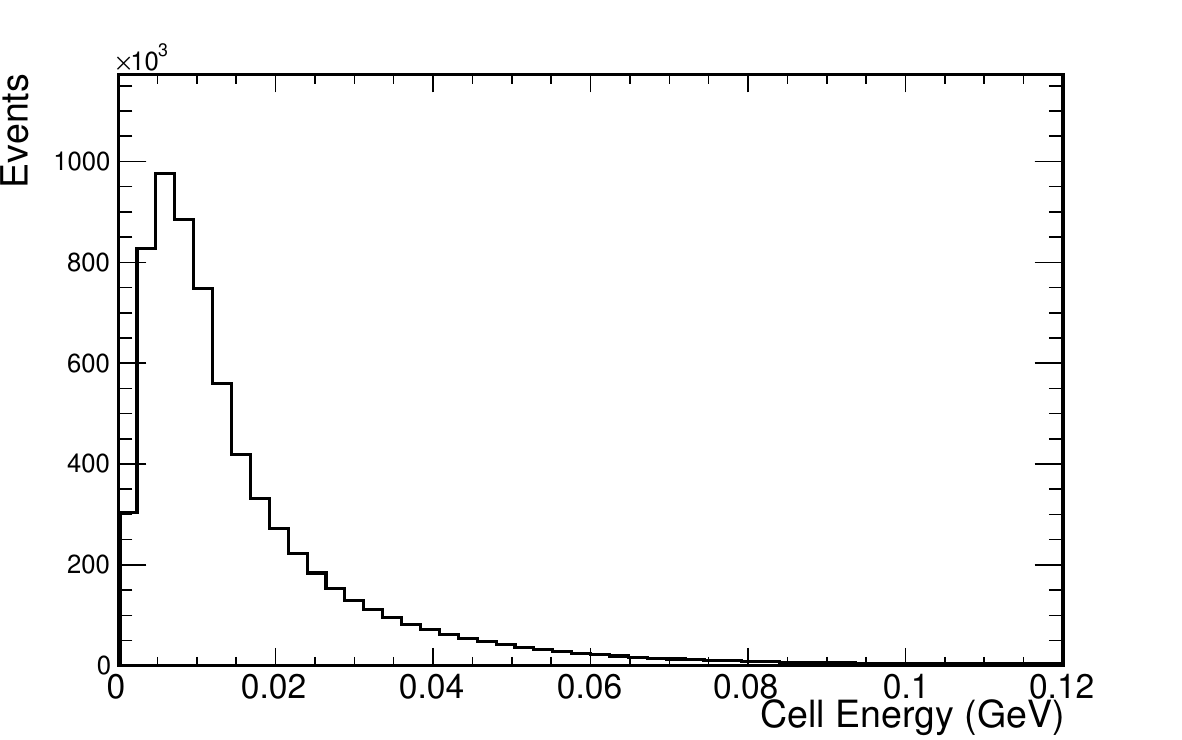}
     \includegraphics[width=8cm]{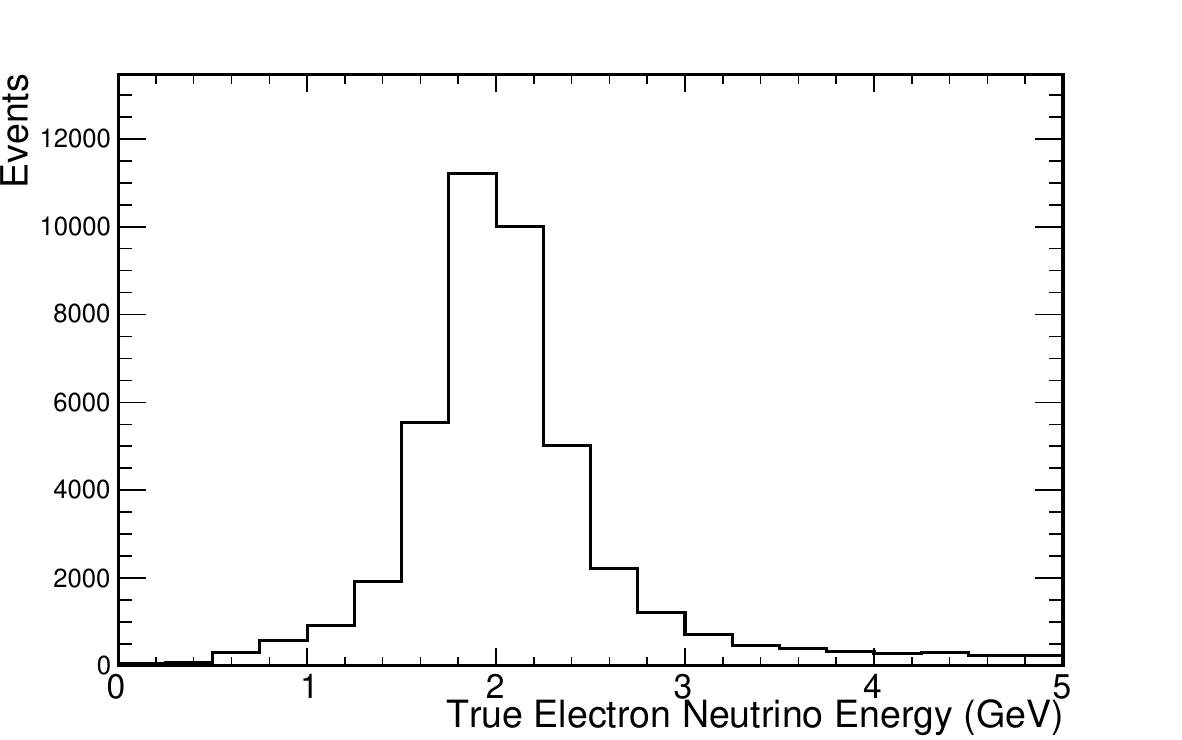}
    \caption{Distributions of input cell energies (left) and true electron neutrino energies  (right) from a subset of the regular flux training sample.}
    \label{fig:nue_trainE_nonflat}
\end{figure}

\subsubsection{The Electron Neutrino Dataset}


True neutrino information is used to select $\nu_e$ CC events from all neutrino interactions. To to speed up the processing time, a loose pre-selection is applied to remove events with long prongs or too many hits. The pre-selection requires the number of occupied cells in the neutrino interaction to be less than 200. Additionally, the length of the longest prong is required to be less than 500 cm. Prongs are collections of cell hits with a start point and direction, which are reconstructed based on distances from hits to the lines associated with each of the particles that paths emanating from the reconstructed vertex~\cite{ref:reco2, ref:reco3, ref:reco4, ref:fuzzyk, Niner:2015aya}. The pre-selection keeps most of the electron-neutrino appearance signal while rejecting a large fraction of background events with long muon tracks. No requirements are applied to calorimetric energy or reconstructed neutrino interaction identities.

We use 0.98 million simulated samples of electron neutrino interactions as the electron neutrino dataset. Each sample consists of a pair of images from the two detector views, the reconstructed vertex and the simulated truth of electron neutrino energy. We split the dataset into a training sample with 0.75 million events and a validation sample with 0.23 million events. One example pair of images from the training dataset is shown in Figure~\ref{fig:nue_image}. In Figure~\ref{fig:nue_trainE_flat} (left) we show the spectrum of cell energy deposits in cell hits from a subset of the flat flux training sample. Figure~\ref{fig:nue_trainE_flat} (right) shows the spectrum of true $\nu_e$ energy from the subset of the flat flux training sample. One can find that there are enough events in the low $\nu_e$ energy region ($<$ 1 GeV) for training. As a comparison, in Figure~\ref{fig:nue_trainE_nonflat} the cell hit energy deposits and $\nu_e$energy from the regular flux $\nu_e$ FD Monte Carlo sample are shown. Since NOvA is an off-axis experiment, the $\nu_e$ energy in the FD is bell-shaped peaking around at $2$ GeV, and there are few events below 1 GeV for training.

\begin{figure}[t]
  \centering
    \begin{minipage}[b]{0.8\linewidth}
    \includegraphics[width=\linewidth]{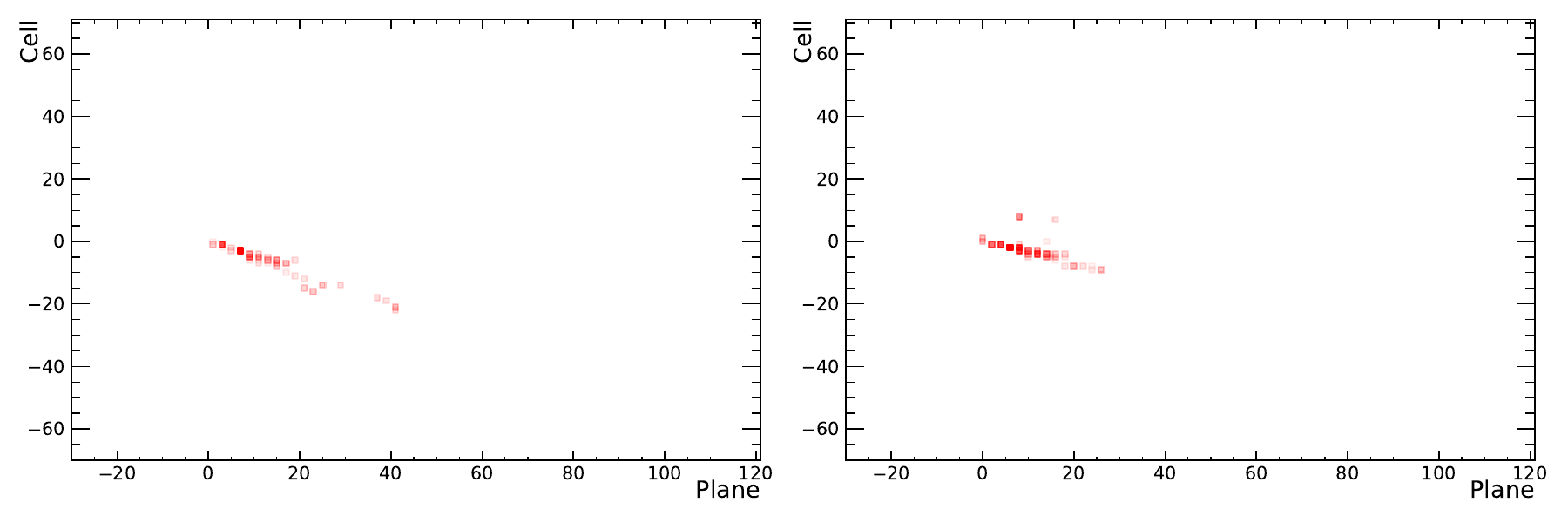}
  \end{minipage}
    \caption{Electron shower image pair example.}
  \label{fig:elec_image}
\end{figure}

 \begin{figure}[t]
    \centering
    \includegraphics[width=8cm]{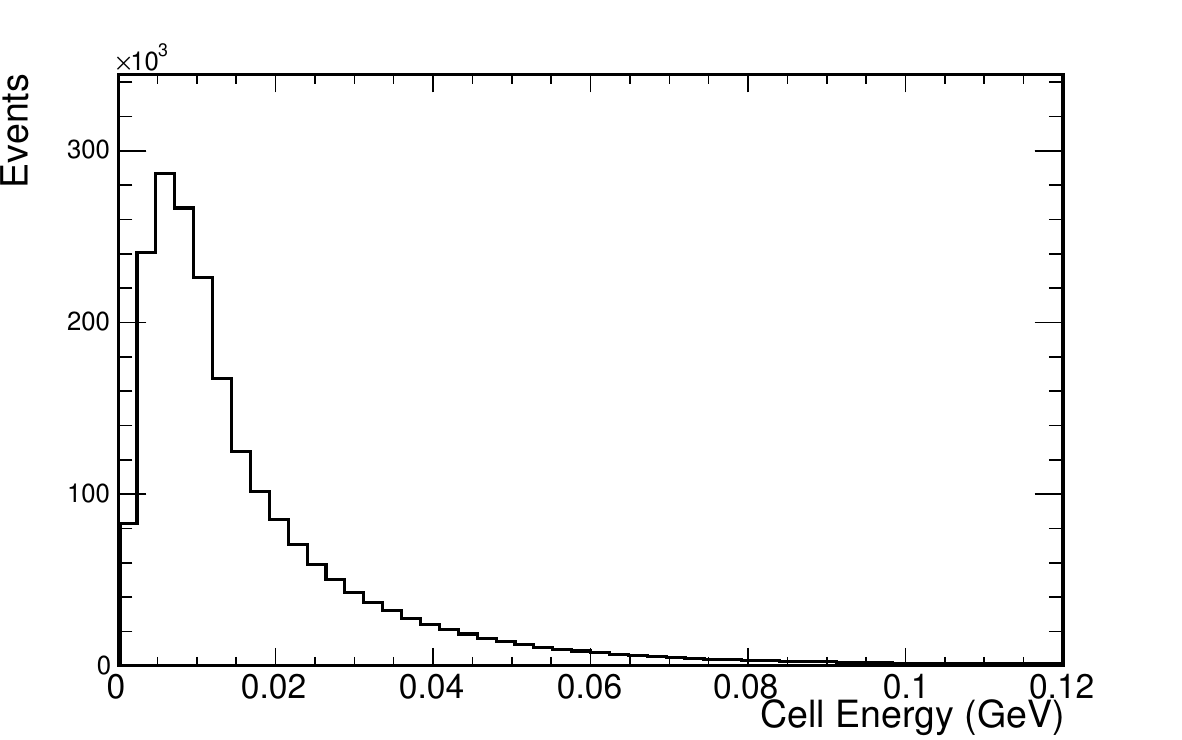}
     \includegraphics[width=8cm]{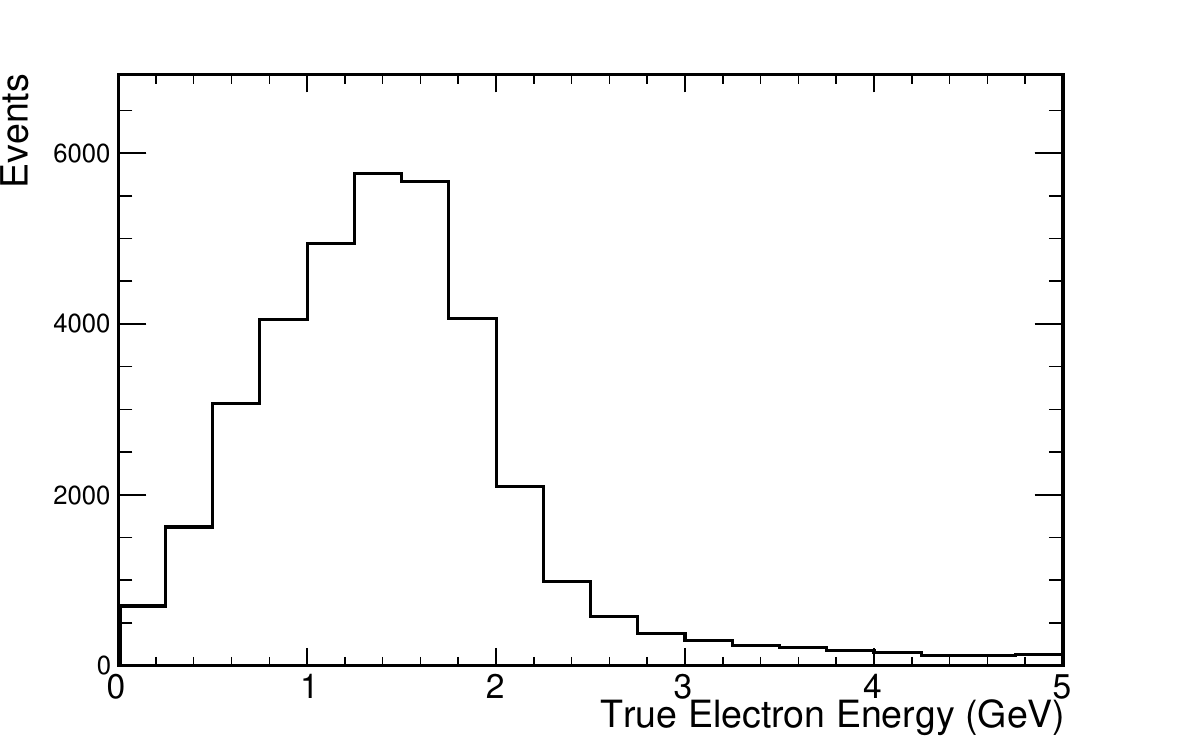}
    \caption{Distributions of input cell energies (left) and true electron energies  (right) from a subset of the regular flux training sample.}
    \label{fig:elec_trainE_nonflat}
\end{figure}

\subsubsection{The Electron Shower Dataset}

Electron shower images are created from electron neutrino interactions by reconstructing the pixels corresponding to the electron shower and setting cell energy values for all other pixels to zero. We select electron showers from $\nu_e$ CC FD Monte Carlo events by matching the reconstructed shower direction to the true particle direction.

The creation of electron shower pixel maps starts with prongs. First, the shower core is defined based on the prong direction provided by the prong cluster. Then signal hits are collected in a column around this core. The electron deposits energy through ionization in the first few planes before it starts multiple scattering.  In order to capture all deposits from the electron shower, the region corresponding to multiple scattering is enlarged. We require the radius to be twice the cell width for the first 8 planes from the start point of the shower. For the following planes, we require the radius to be $20$ times the cell width~\cite{Bian:2015opa}.

Electrons from the regular flux $\nu_e$ CC FD Monte Carlo sample are used for training. One example pair of images from the training dataset is shown in Figure~\ref{fig:elec_image}. We use 660k simulated samples of electron showers as the electron shower dataset. Each sample consists of a pair of images, the reconstructed vertex and the true electron energy. We split the dataset into 610,000 training samples and 50,000 validation samples.  In Figure~\ref{fig:elec_trainE_nonflat} (left) we show the spectrum of cell energy deposits from a subset of the regular flux training sample. Figure~\ref{fig:elec_trainE_nonflat} (right) shows the spectrum of true electron shower energies from the subset of training sample.

\subsection{Neural Network Architecture}\label{sec:neural-network-architecture}

\begin{figure*}[h]
    \centering
    \includegraphics[width=.8\linewidth]{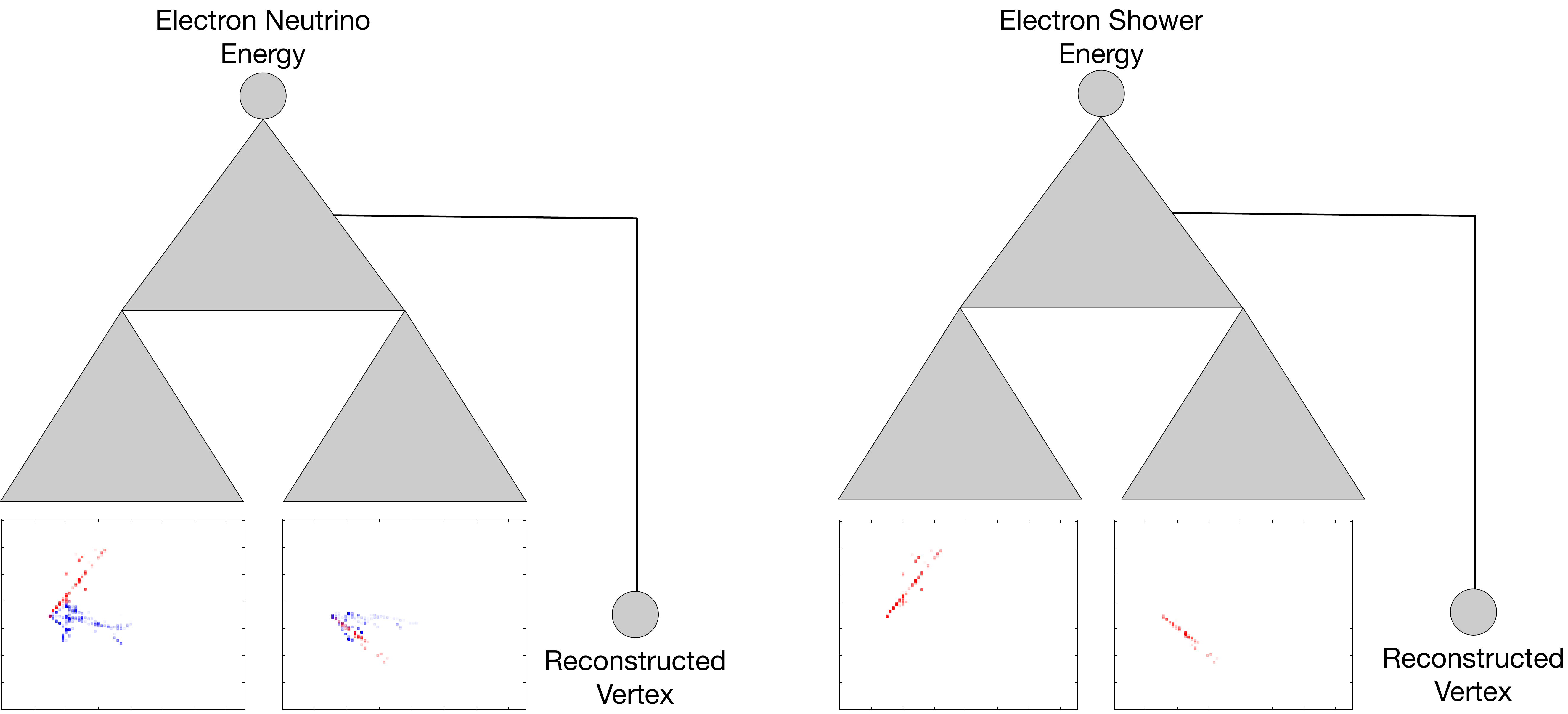}
    \caption{Diagram of electron neutrino energy predictor (left) and electron shower energy predictor (right). Triangles represent neural networks.}
    \label{fig:two_predictors}
\end{figure*}

The electron neutrino energy model and electron shower energy model are equal in architecture but weights are not shared across the two predictors. The neural network input consists of two matrices of shape $(151, 141)$ which represent the pixel values of the images and the cell indices for each view given by the reconstructed vertex. The output is one positive real-valued number. Inputs and outputs for each model are illustrated in Figure~\ref{fig:two_predictors}.

The neural network architecture is modified from the architecture used by NOvA's CVN event classifier~\cite{Aurisano:2016jvx}. This architecture is optimized from the convolutional neural network GoogLeNet~\cite{ref:googlenet} developed for image recognition. In addition, NOvA's CVN utilizes a siamese network structure \cite{baldi93finger}. The siamese network structure consists of two identical sub-networks whose outputs are merged to produce the final output. Each sub-network processes an image from one view. Weights are not shared between the sub-networks to provide independent information aggregation in each view. The sub-networks are constructed from convolutional layers and pooling layers.

Convolutional layers apply a weight matrix in a sliding window fashion to the input image. This allows computing the same feature at different locations in the input image producing an output referred to as a feature map. Using multiple weight matrices allows learning a variety of features from the input images. Stacking these convolutional layers makes it possible to learn higher level features with each additional layer. Pooling layers take a feature map and reduce its dimensionality. This is done by tiling the feature map and reducing each tile to its maximum or average. Pooling layers are normally used between convolution layers. The convolutional layers in our network are based on the Inception module introduced by Ref.~\cite{ref:googlenet} as part of GoogLeNet.

GoogLeNet is the winner of the ImageNet Large Scale Visual Recognition Competition 2014 \cite{russakovsky2015imagenet} and state-of-the-art for convolutional neural networks with pixelmap inputs. Therefore, it is a good choice for our task. In particular, GoogLeNet uses the Inception module to efficiently extract features of different sizes from the input. This increases its modeling capacity without a significant increase in the computational cost.

Each sub-network here is constructed from a sequence of \texttt{Conv}-\texttt{MaxPool}-\texttt{Conv}-\texttt{Conv}-\texttt{MaxPool} layers followed by two Inception modules \cite{ref:googlenet}. The Inception module is a specific configuration of convolutional layers built to simultaneously extract features of different dimensions. The features are then concatenated and pooled. Each convolutional layer and Inception module has $32$ filters in the networks used. In the experiments, using additional Inception modules does not improve model performance. This is expected because the images here are sparse relative to natural images.

In the NOvA far detector, the scintillation light produced by neutrino interactions in each 15.5m-long detector cell is collected by the wavelength shifting fiber and read from the avalanche photodiode installed on one end of the cell. The attenuation of the scintillation light signal is a function of the distance from the interaction point to the readout photodiode, so the number of photoelectrons on the photodiode depends on the location of the interaction point. The readout threshold for each cell is a fixed number of photoelectrons, so the distribution of the cell energy deposit in each cell after the readout threshold cut is impacted by the position of the interaction with respect to the readout. This position dependence cannot be recovered by the cosmic attenuation calibration, which corrects the position dependence of the average number of photoelectrons for each cell using the minimum ionizing peak (MIP) position of cosmic muon hits. To consider this position effect in cell energies, we use the reconstructed vertex positions in the two views as neural network inputs. In our neural network architecture, after the inception modules and an average pooling layer, the output is flattened and concatenated with the reconstructed vertex position. A linear regression follows which produces the network output, the electron neutrino energy.

\subsection{Neural Network Training}\label{sec:neural-network-training}

Neural networks for supervised learning are trained by defining a differentiable loss function $L$ between neural network outputs $f_{\mathbf{W}}(\mathbf{x}_i)$ and target values $y_i$. Here, $\mathbf{W}$ represents the weights of the neural network and $\mathbf{x}_i$ is the neural network input. The loss function represents the metric by which the neural network accuracy is assessed. During training the neural network weights $\mathbf{W}$ are iteratively updated to minimize $L$ using its gradient and a step size $\alpha$.

Typically these updates are computed over mini-batches of size $n$ which make up a partition of the total training dataset. Iteration over all mini-batches constitutes one epoch. Training parameters such as the step size  $\alpha$ and the batch size $n$ are referred to as hyperparameters must be selected before training and possibly tuned using the loss function on the validation dataset. We utilize the hyperparameter optimization software SHERPA \cite{ref:sherpa2018} which implements a number of hyperparameter optimization strategies and visualization tools. By automating the task, this software significantly speeds up the computationally expensive of finding optimal hyperparameters.

Unlike classification problems such as image recognition and particle identification, the target value $y_i$ for the output (energy) of our regression neural network is a continuous variable varying over events. This requires the definition of an appropriate loss function for the task of the regression neural network.
The goal is to minimize the standard deviation of a Gaussian fit to the peak of the histogram given by the energy resolution $\frac{E_{reco}-E_{true}}{E_{true}}$ on the test set. While this quantity cannot be directly optimized the absolute scaled error loss provides an appropriate surrogate for the task. The training loss function is then given by:
\begin{equation}
    L(\mathbf{W}, \{\mathbf{x}_i, y_i\}_{i=1}^n ) = \frac{1}{n} \sum_{i=1}^n |\frac{f_{\mathbf{W}}(\mathbf{x}_i)-y_i}{y_i}|.
\end{equation}

Traditional loss functions for regression problems are the mean squared error $\frac{1}{n} \sum_{i=1}^n (f_{\mathbf{W}}(\mathbf{x}_i)-y_i)^2$ or the mean absolute error $\frac{1}{n} \sum_{i=1}^n |f_{\mathbf{W}}(\mathbf{x}_i)-y_i|$. The former is often used due to its relationship to the log-likelihood when the data distribution is assumed to be Normal, its strict convexity, the fact it can be decomposed into variance and bias, and many other desirable properties. In our case, however, the mean squared error is suboptimal, because its derivative with respect to $\mathbf{W}$ is $\frac{1}{n} \sum_{i=1}^n 2 (f_{\mathbf{W}}(\mathbf{x}_i)-y_i) \frac{\delta f_{\mathbf{W}}(\mathbf{x}_i)}{\delta \mathbf{W}}$. This increases proportionally with the distance of the predicted value from the truth. In other words, outliers will have increased impacts during gradient descent. In the training for neutrino energy, the events with large invisible energy due to dead material and hadronic interactions shouldn't have much larger impacts than those whose visible energy is close to the true energy, so we choose the absolute error instead of the squared error in the loss function. Furthermore, the original CVN/GoogleNet are designed and trained for classification tasks, we optimized training hyperparameters for the regression task.

Input image pixels are typically normalized to increase numerical stability and gradient quality. Here most image pixels are zero and the non-zero ones tend to be small. We apply three normalization methods: mean zero unit variance standardization, log transformation, and constant scaling. The three methods produce similar results. Therefore, a constant scaling factor of $100$ is chosen after visual inspection of the input spectrum for $\nu_e$ (Figure~\ref{fig:nue_trainE_nonflat} (right)) and electron (Figure~\ref{fig:elec_trainE_nonflat} (right)).

The models are trained with stochastic gradient descent. The hyperparameter search yields as best hyperparameters: (1) a batch size of $n=32$; (2) an initial learning rate of $5\times10^{-4}$ with an exponential learning rate decay per batch of $1\times10^{-5}$; and (3) a momentum of $0.7$. Models are trained for 100 epochs, or until the validation loss does not increase by at least 0.001 for 5 epochs. The weights from the epoch with the best validation loss are kept.



Regularization techniques are also explored using random search implemented in SHERPA. Regularization refers to a set of methods that reduce modeling capacity to prevent fitting to noise in the training set (over-fitting). Here, the model training is optionally regularized with L2-penalty on all convolutional layer weights and on the fully connected layer weights. L2-penalty also referred to as weight-decay which adds the term $\lambda ||\mathbf{W}||_2^2$ to the loss function $L(\mathbf{W}, \{\mathbf{x}_i, y_i\}_{i=1}^n )$. The added term prevents weights from getting too large and thus reduces modeling capacity. Random search was applied to the L2-penalty multiplier $\lambda$ over a range of $1\times10^{-5}$ to $1\times10^{-7}$.  To increase the robustness of learned features, dropout \cite{ref:srivastava2014dropout, ref:baldidropout14} was also applied to the fully-connected layer. Dropout is a technique that randomly sets hidden layer units to zero with a given dropout-probability during the training. In the hyperparameter search, we let the dropout-probabilities range from $0$ to $0.4$. The best performing model found from random search had $\lambda=0$ and zero dropout-probability. While dropout and L2-penalty tend to be useful for classification there is an intuitive explanation as to why those methods decrease performance in our regression problem. In the case of classification, outputs do not directly depend on the magnitude of the neural network outputs since the outputs are normalized by the sum of all outputs. In regression, the output of the neural network has to exactly match the target value, which can take a wide range of values. If hidden layer units are randomly dropped as in dropout, this estimate may significantly change depending on what units are dropped. Similarly, L2-penalty may prevent weights from adopting the magnitude required by the scale of the targets of the prediction.

For the training and validation of the neural network, all models are implemented in Keras \cite{chollet2015keras} with Tensorflow backend.

\section{Results}\label{sec:result}

We use a simulated test data sample independent of the training and validation samples in Section~\ref{sec:method} to test the physics performance of the trained neural networks. The test $\nu_e$ CC sample is produced with the same simulation and reconstruction method as the training sample. Simulation and reconstruction at NOvA are described in Section~\ref{sec:sample}.  The test sample has a regular flux. In order to mimic the real neutrino energy spectrum in the NOvA FD, we apply the neutrino oscillations to each FD MC sample by event weighting. To mimic overall energy resolution of the $\nu_e$ oscillation signal while keeping independent from specific CP and mass order choices, oscillation probabilities are calculated from the first-order terms in the full oscillation formula. The  values of oscillation parameters are chosen to be $\sin^2\theta_{23} = 0.5$, $\sin^22\theta_{23} = 1$, $\Delta m^2_{32} = +2.35\times10^{-3}$eV$^{2}$ and $\sin^{2}2\theta_{13} = 0.1$.

\subsection{$\nu_e$-CC neutrino energy}

The proposed regression CNN energy estimator is compared with two methods used in previous NOvA $\nu_e$ analyses: calorimetric energy estimation and kinematics-based energy estimation. Used as the $\nu_e$ CC energy in NOvA's first $\nu_e$ oscillation analysis in 2016~\cite{Adamson:2016tbq}, the calorimetric energy estimator takes the sum of the calibrated calorimetric energy in each cell for an event and multiplies the sum by a scale factor. The scale factor corrects for the dead material in NOvA detectors and missing energy taken by undetected particles. It is estimated via simulated neutrino events. The kinematics-based energy estimator is based on the method used in NOvA's $\nu_e$ analysis in 2017~\cite{ref:taenergy} (Kinematic Energy). This estimator is based on a quadratic function of the reconstructed electromagnetic and hadronic energy. The electromagnetic energy component is estimated by the sum of calorimetric energies from the electron and photons. The hadronic energy is estimated via the sum of calorimetric energies from hadrons such as pions, kaons, and protons. The electron, photons, and hadrons are identified by a deep-learning based particle identification algorithm called prong CVN \cite{Psihas:2018czu}.  Parameters of the quadratic function are also determined using simulated data.

\begin{figure}[h]
    \centering
    \includegraphics[width=8cm]{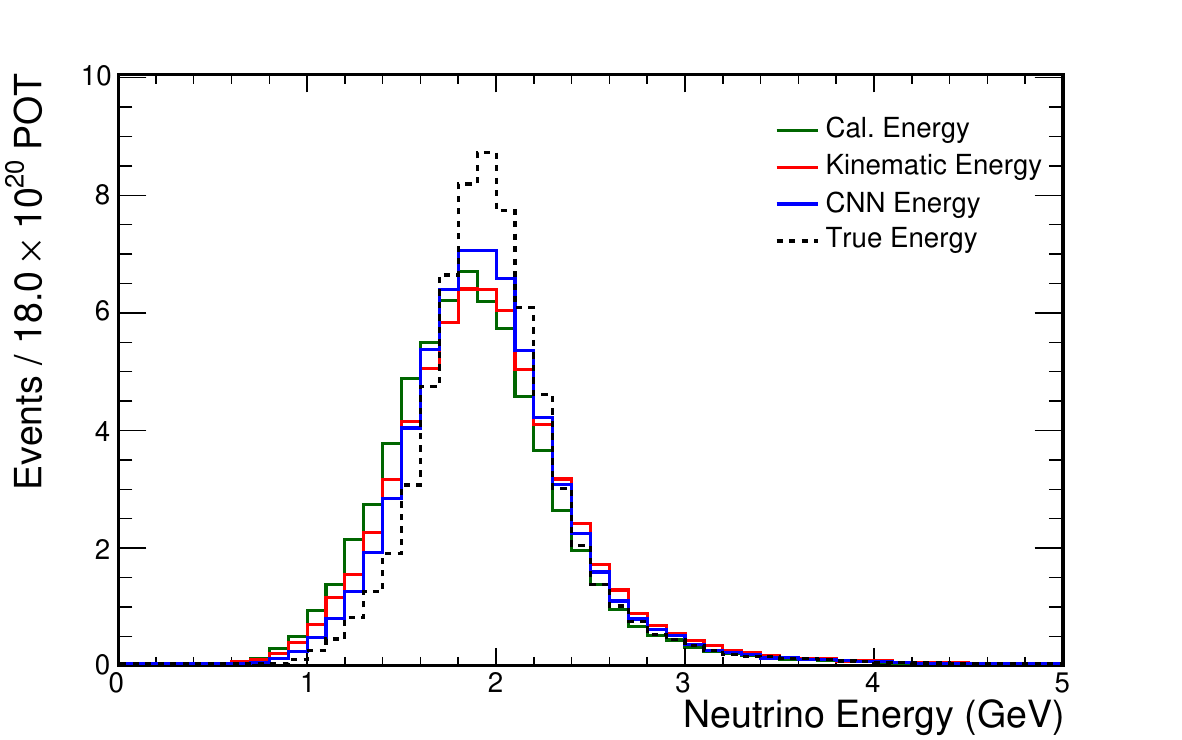}
    \caption{Monte Carlo distributions of $\nu_e$ CC energy reconstructed by the regression CNN (CNN Energy, blue), particle kinematic information (Kinematic Energy, red), and summing the calibrated calorimetric energy in each cell (Calorimetric Energy, green), overlapping with true neutrino energy (dashed). The neutrino oscillations are applied.}
    \label{fig:nueE}
\end{figure}

The true and reconstructed $\nu_e$ CC energy in the FD, weighted by the oscillation probabilities, are shown in Figure \ref{fig:nueE}. The off-axis spectrum convoluted with the oscillation probability makes the $\nu_e$-CC energy spectrum peak at around 2 GeV. 

\begin{figure}[h]
    \centering
    \includegraphics[width=8cm]{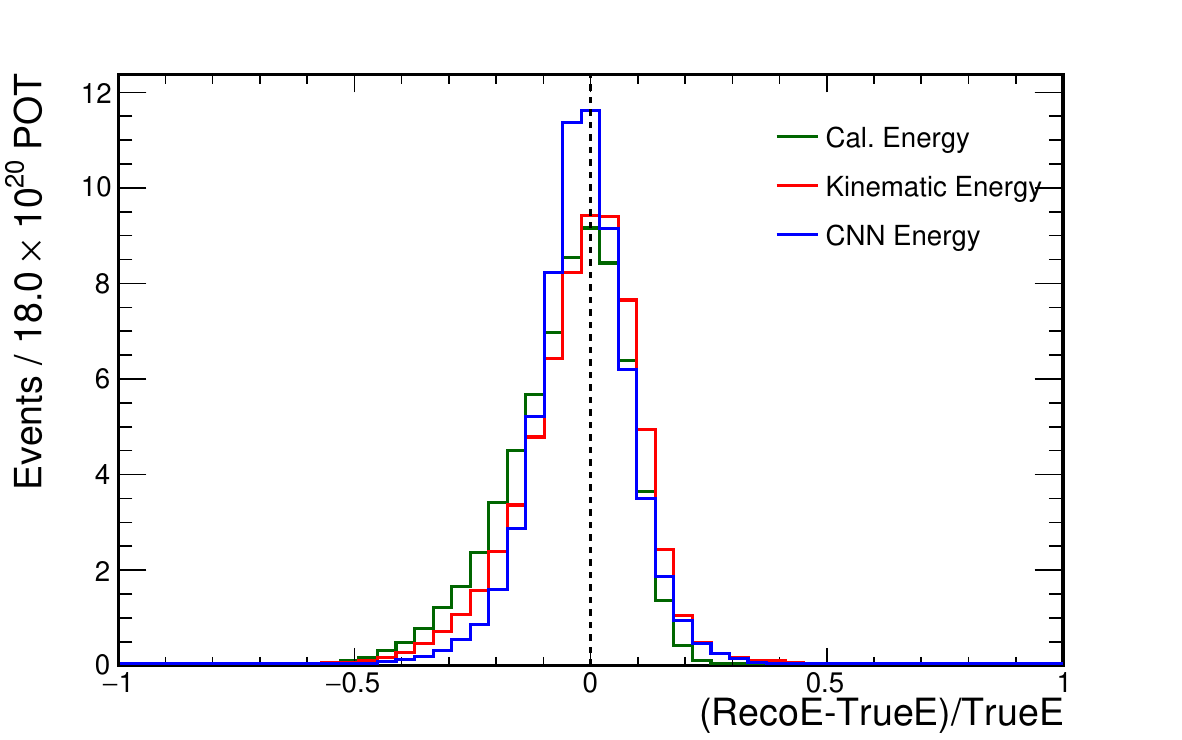}
    \caption{Monte Carlo distributions of ratios of differences between reconstructed and true $\nu_e$ CC energies to true $\nu_e$ CC energy in the calorimetric energy range of 0 to 5 GeV. Neutrino energy is reconstructed by the regression CNN (CNN Energy, blue), particle kinematic information (Kinematic Energy, red), and summing the calibrated calorimetric energy in each cell (Calorimetric Energy, green). The neutrino oscillations are applied.}\label{fig:nueERes}
\end{figure}

The overall performance is illustrated in Figure~\ref{fig:nueERes}. Shown are histograms of ($E_{reco}-E_{true})/E_{true}$, the ratio of the difference between reconstructed and true $\nu_e$ CC energy over true neutrino energy in the calorimetric energy range of 0 to 5 GeV. The neural network energy is the one with the best resolution. Gaussian fits to $(E_{reco}-E_{true})/E_{true}$ distributions provide relative resolutions of $8.9\%$ (CNN Energy), $10.1\%$ (Kinematic Energy) and $10.2\%$ (Calorimetric Energy), respectively. The relative resolution is defined as the ratio of the standard deviation to the peak value in a Gaussian fit. Relative RMSs (the ratio of RMS to Mean) are $11.1\%$ (CNN Energy), $13.2\%$ (Kinematic Energy) and $13.6\%$ (Calorimetric Energy). 

\begin{figure}[h]
    \centering
    \includegraphics[width=8cm]{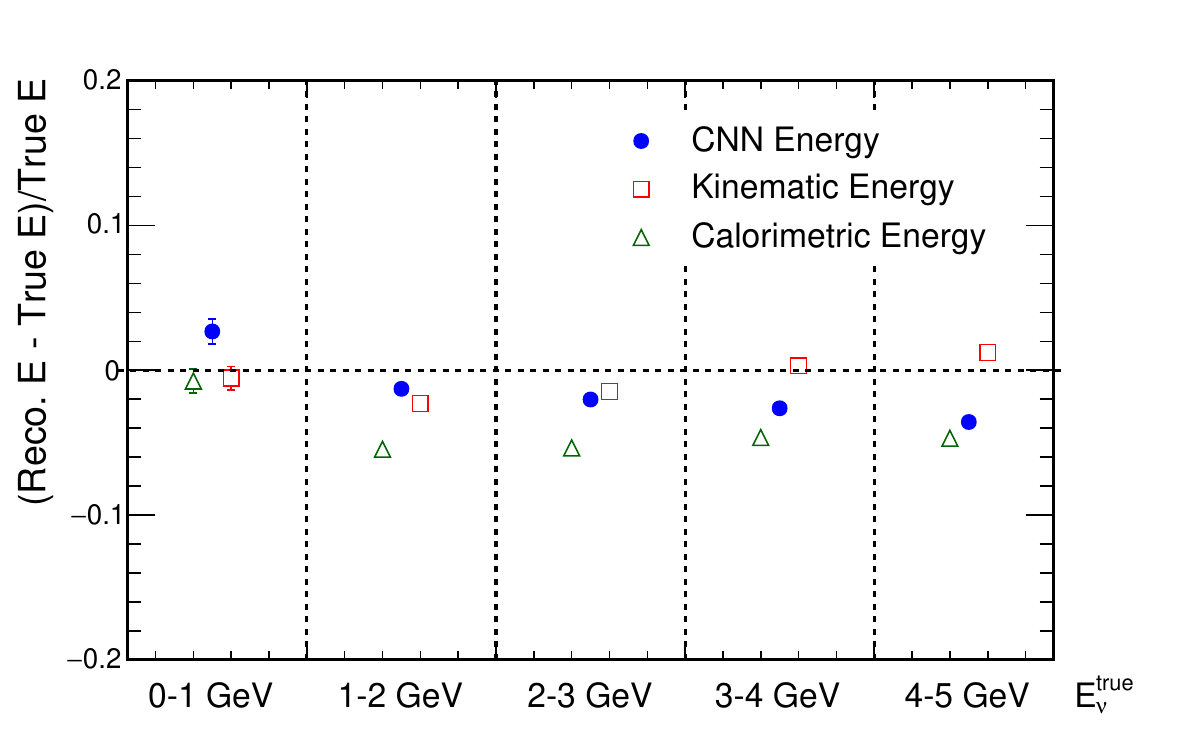}
     \includegraphics[width=8cm]{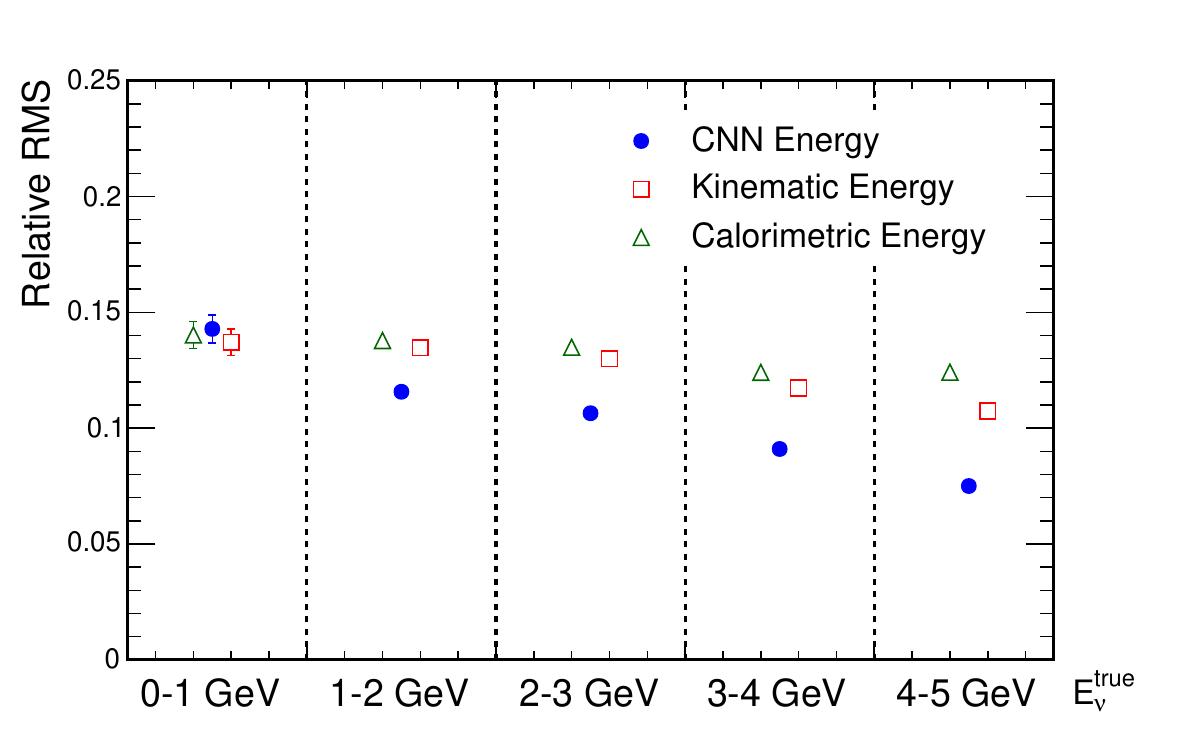}
    \caption{Means (left) and relative RMS (right, the ratio of RMS to the mean value of energy) of the Monte Carlo distributions of the ratios of differences between reconstructed and true $\nu_e$ CC energies to true $\nu_e$ CC energy for different true neutrino energy bins ranging from 0 to 5 GeV. Neutrino energy is reconstructed by the regression CNN (CNN Energy, blue), particle kinematic information (Kinematic Energy, red), and summing the calibrated calorimetric energy in each cell (Calorimetric Energy, green). The neutrino oscillations are applied.}
    \label{fig:nueEResvsTrueE}
\end{figure}

Figure~\ref{fig:nueEResvsTrueE} shows means and RMSs of $(E_{reco}-E_{true})/E_{true}$ in each 1-GeV-wide true energy bin. Both Kinematic Energy and CNN energy estimators are determined from the NOvA FD $\nu_e$ CC signal sample with oscillated energy spectrum, peaking around 2 GeV. As shown in Figure~\ref{fig:nueEResvsTrueE} (left), the energy scale of the three estimators shows no significant biases with respect to the true neutrino energy, and the regression CNN has better energy resolutions.

The standard training sample of the described regression CNN $\nu_e$ energy estimator uses a flat flux. We also train the regression CNN using the regular flux with the peak around 2 GeV to understand the effect of the training energy spectrum on the linearity of the energy scale. The flat flux sample and the regular flux sample are defined in Section~\ref{sec:neural-network-training}.

\begin{figure}[t]
    \centering
    \includegraphics[width=8cm]{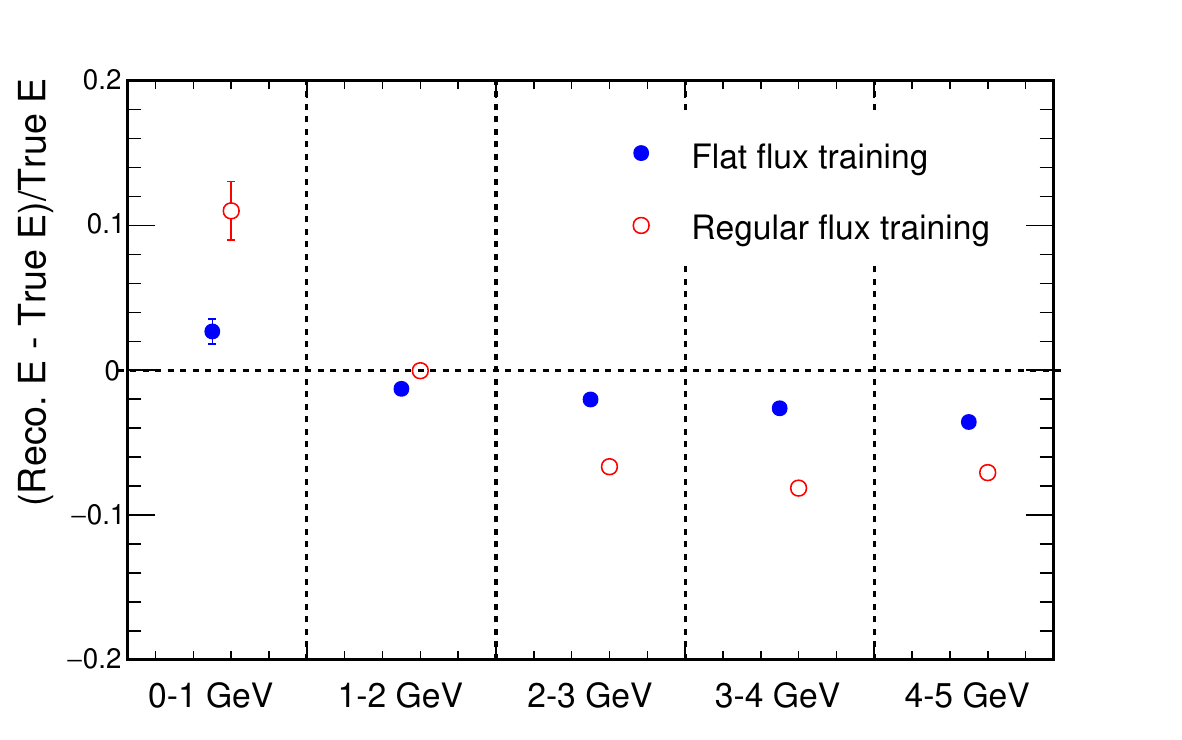}
    \caption{Means of the Monte Carlo distributions of ratios of differences between reconstructed and true $\nu_e$ CC energies to true $\nu_e$ CC energy for different true neutrino energy bins ranging from 0 to 5 GeV. Neutrino energy is reconstructed by CNN trained with flat flux (blue) and regular flux (red), the neutrino oscillations are applied.}
    \label{fig:nueEResvsTrueE0}
\end{figure}

Energy scales for neutrino energy based on the flat flux training and regular flux training are shown in Figure~\ref{fig:nueEResvsTrueE0}. One can find that the energy scale from the flat flux training has less biases over true neutrino energy. The flat flux training, therefore, represents the preferred training mode to generate the regression CNN for the neutrino energy reconstruction.

\begin{figure}[t]
    \centering
    \includegraphics[width=16cm]{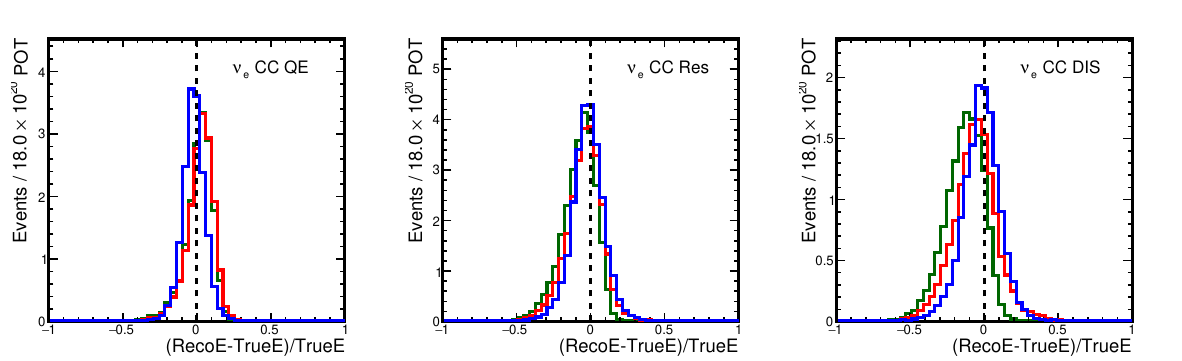}
    \caption{Monte Carlo distributions of ratios of the difference between reconstructed and true $\nu_e$ CC energy to true neutrino energy for QE, RES and DIS modes. Neutrino energy is reconstructed by the regression CNN (CNN Energy, blue),  particle kinematic information (Kinematic Energy, red), and summing the calibrated calorimetric energy in each cell (Calorimetric Energy, green). The neutrino oscillations are applied.}
    \label{fig:nueEResMode}
\end{figure}

Figure~\ref{fig:nueEResMode} shows estimator performance by interaction mode. $\nu_e$ CC interactions can be classified as quasi-elastic (QE), resonant (RES), and deep-inelastic scattering (DIS) modes. In a QE event, the nucleon ($p$ or $n$) recoils quasi-elastically from the scattering electron, and the electron, because of its small mass, takes the majority of the incident neutrino energy.  Hadronic energy portions and hadron multiplicities vary in these three modes. For the RES mode, the nucleon is excited into baryonic resonances and decays to hadrons, so more neutrino energy is transferred into the hadronic system. In DIS events, the nucleon is smashed into several hadrons, requiring even larger neutrino energy transfer to the hadronic system. Figure~\ref{fig:nueEResMode} shows $(E_{reco}-E_{true})/E_{true}$ in these categories individually. The CNN Energy scale shows a better resolution and consistency among the interaction modes.

\begin{figure}[h]
    \centering
    \includegraphics[width=8cm]{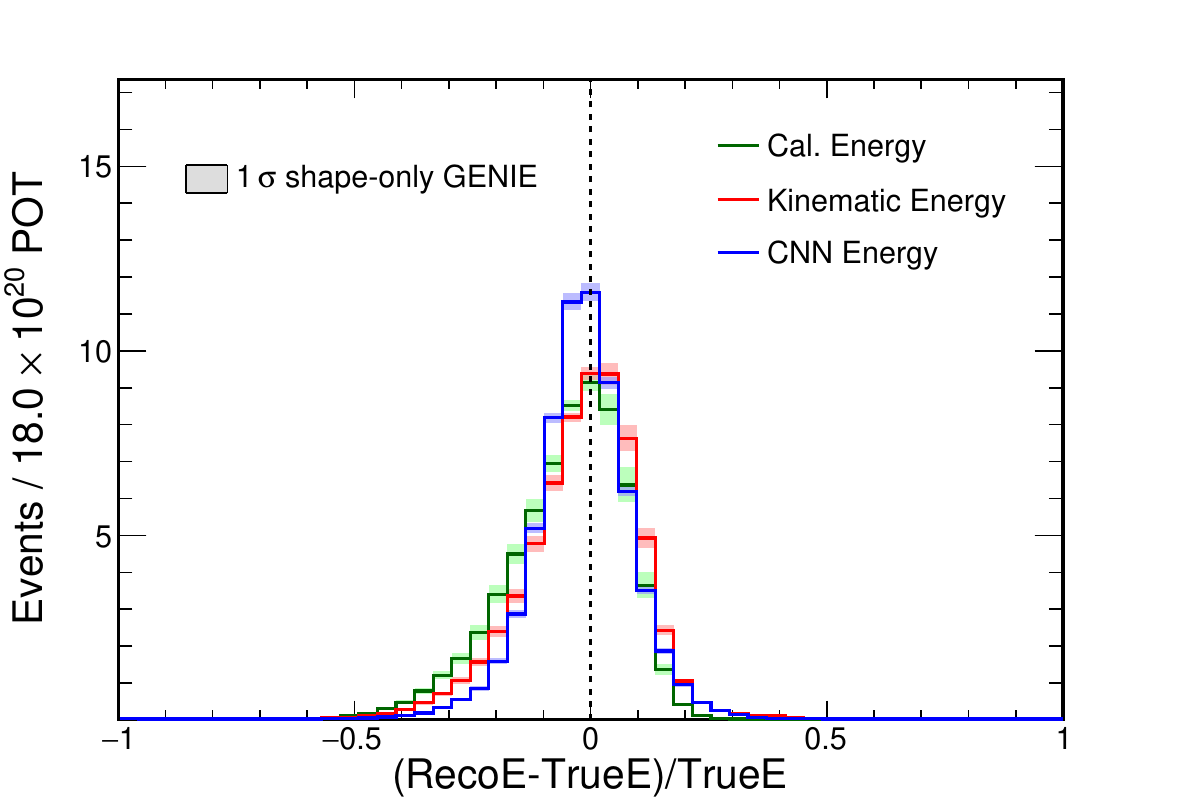}
    \caption{Monte Carlo distributions of ratios of differences between reconstructed and true $\nu_e$ CC energies to true $\nu_e$ CC energy in the calorimetric energy range of 0 to 5 GeV. Error bands represent systematic uncertainties evaluated by GENIE reweighting. Neutrino energy is reconstructed by the regression CNN (CNN Energy, blue), particle kinematic information (Kinematic Energy, red), and summing the calibrated calorimetric energy in each cell (Calorimetric Energy, green). The neutrino oscillations are applied.}
    \label{fig:res_1d_syst}
\end{figure}

Systematic uncertainties in the energy reconstruction from the simulation of neutrino interactions are evaluated by using the reweighting knobs built into GENIE~\cite{Andreopoulos:2015wxa}. Each reweighting knob computes a weighting factor that can be applied to MC events to vary normalization and/or shape of a specific type of interaction. In general, these GENIE reweighting knobs deal with systematic uncertainties from modeling of cross-sections, the hadronization, and final state interactions. The reweighting knobs used in this GENIE uncertainty study are similar to NOvA's oscillation analysis Ref~\cite{NOvA:2018gge}. We vary each reweighting knob by +1 $\sigma$ and -1 $\sigma$, where the size of the systematic variation $\sigma$ is the recommendation from the GENIE and NOvA authors, based on surveys of interaction models and existing experimental results. Both the background yield in the signal region before the background correction and the background correction factor determined by the Data-MC difference in the sideband are re-determined in the reweighted background MC. 

\begin{figure}[h]
    \centering
    \includegraphics[width=8cm]{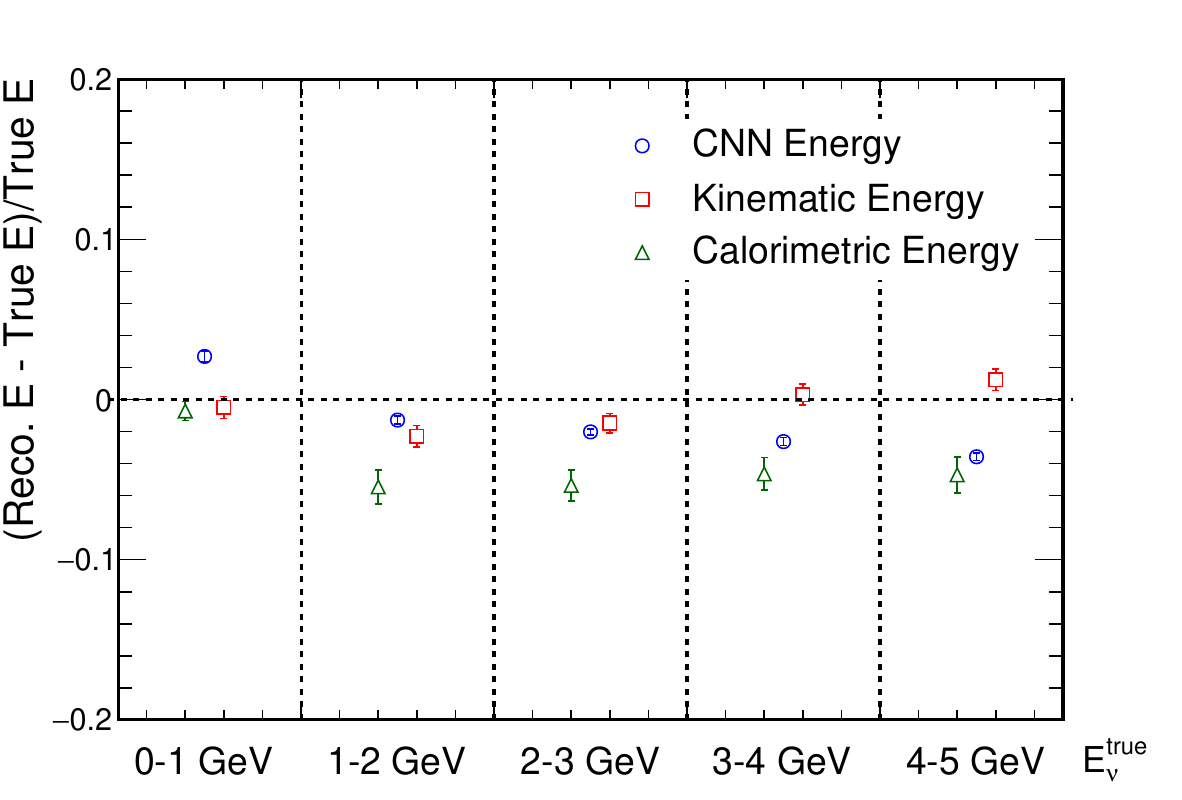}
     \includegraphics[width=8cm]{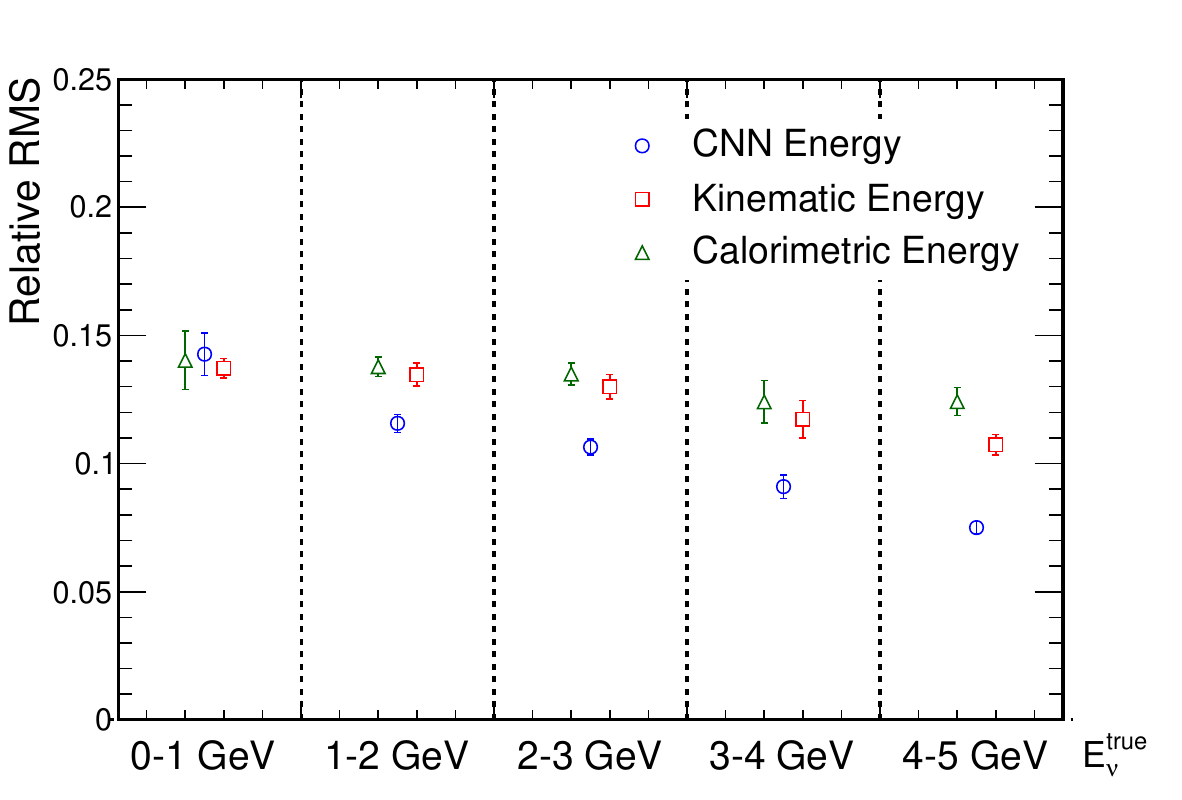}
    \caption{Means (left) and relative RMS (right) of the Monte Carlo distributions of the ratios of differences between reconstructed and true $\nu_e$ CC energies to true $\nu_e$ CC energy for different true neutrino energy bins ranging from 0 to 5 GeV. Error bars represent systematic uncertainties evaluated by GENIE reweighting. Neutrino energy is reconstructed by the regression CNN (CNN Energy, blue), particle kinematic information (Kinematic Energy, red), and summing the calibrated calorimetric energy in each cell (Calorimetric Energy, green). The neutrino oscillations are applied.}
    \label{fig:mean_rms_syst}
\end{figure}

The overall performance with GENIE systematic shifts is illustrated in Figure~\ref{fig:res_1d_syst}. Shown are histograms of $(E_{reco}-E_{true})/E_{true}$, the ratio of the difference between reconstructed and true $\nu_e$ CC energy over true neutrino energy in the calorimetric energy range of 0 to 5 GeV. The systematic errors of $(E_{reco}-E_{true})/E_{true}$ are $0.2\%$ (CNN Energy), $0.6\%$ (Kinematic Energy) and $0.9\%$ (Calorimetric Energy), respectively. The systematic errors of the relative RMSs are $0.3\%$ (CNN Energy), $0.4\%$ (Kinematic Energy) and $0.4\%$ (Calorimetric Energy). Systematic errors of mean and RMS in each energy bin are shown in \ref{fig:mean_rms_syst}. The regression CNN shows smallest systematic uncertainties from the simulation of neutrino interactions.

\subsection{Electron Shower Energy}

\begin{figure}[h]
    \centering
    \includegraphics[width=8cm]{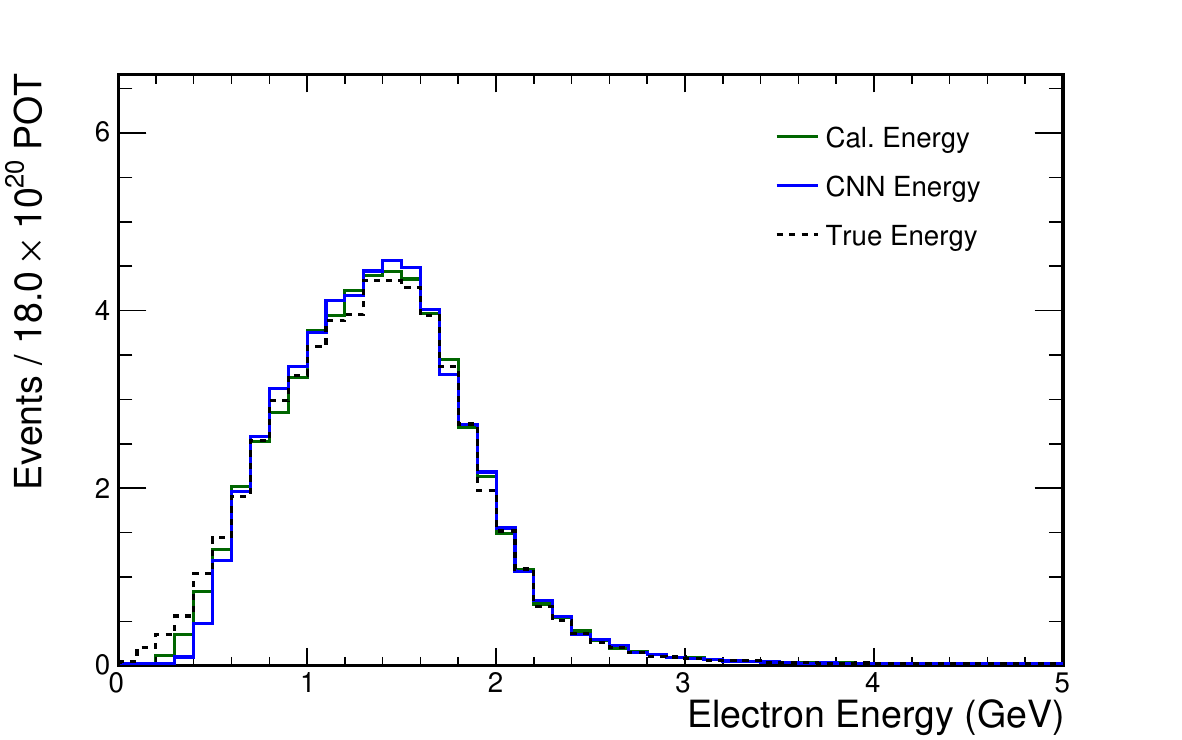}
    \caption{Monte Carlo distributions of reconstructed electron shower energy and true electron energy (dashed) in $\nu_e$ CC events. The shower energy is reconstructed by CNN (CNN Energy, blue) and summing the calibrated calorimetric energy in each cell (Calorimetric Energy, green). The neutrino oscillations are applied.}
    \label{fig:elecE}
\end{figure}

\begin{figure}[h]
    \centering
    \includegraphics[width=8cm]{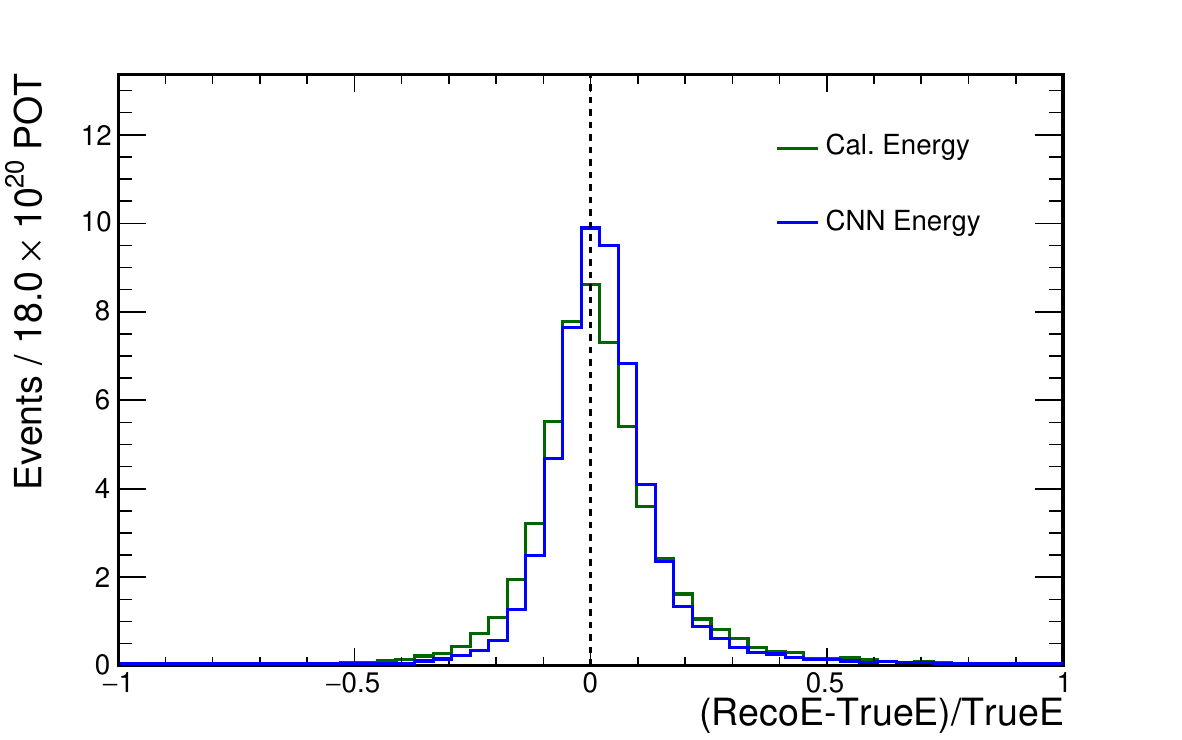}
    \caption{Monte Carlo distributions of ratios of reconstructed electron shower energy to true electron energy in $\nu_e$ CC events in the shower calorimetric energy range of 0 to 5 GeV. Shower energy is reconstructed by CNN (CNN Energy, blue) and summing the calibrated calorimetric energy in each cell (Calorimetric Energy, green). The neutrino oscillations are applied.}
    \label{fig:elecERes}
\end{figure}

\begin{figure}[h]
    \centering
    \includegraphics[width=8cm]{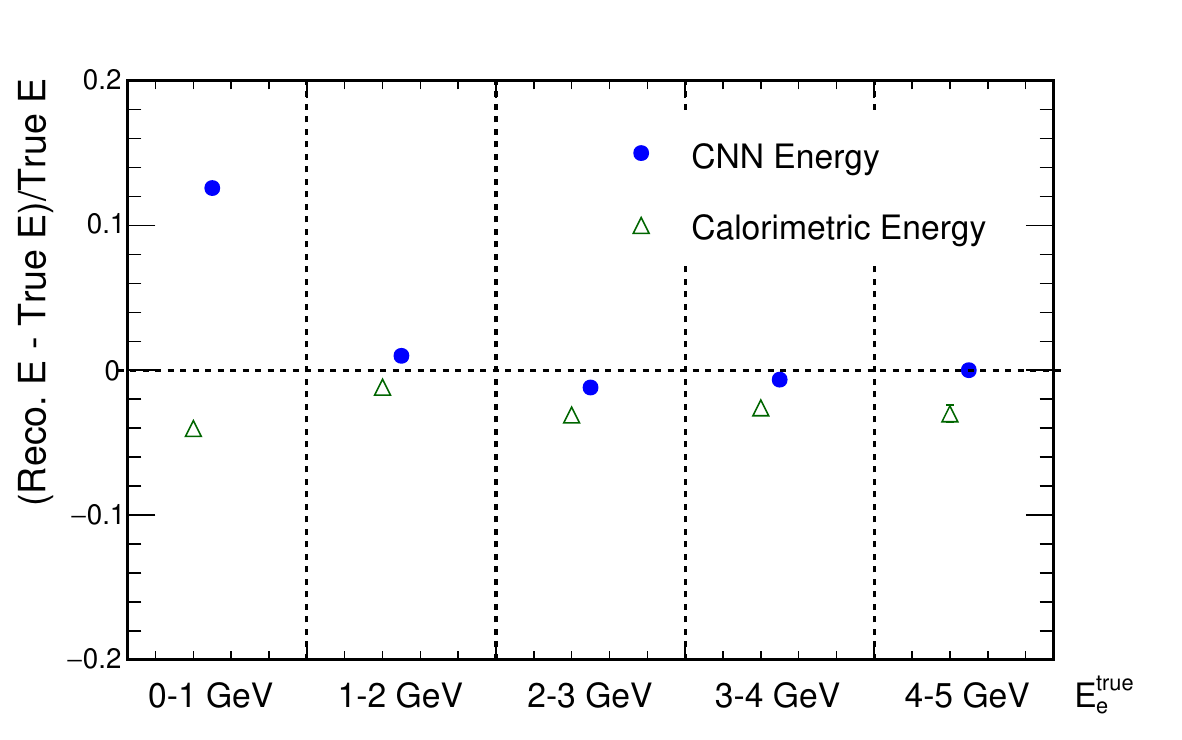}
    \includegraphics[width=8cm]{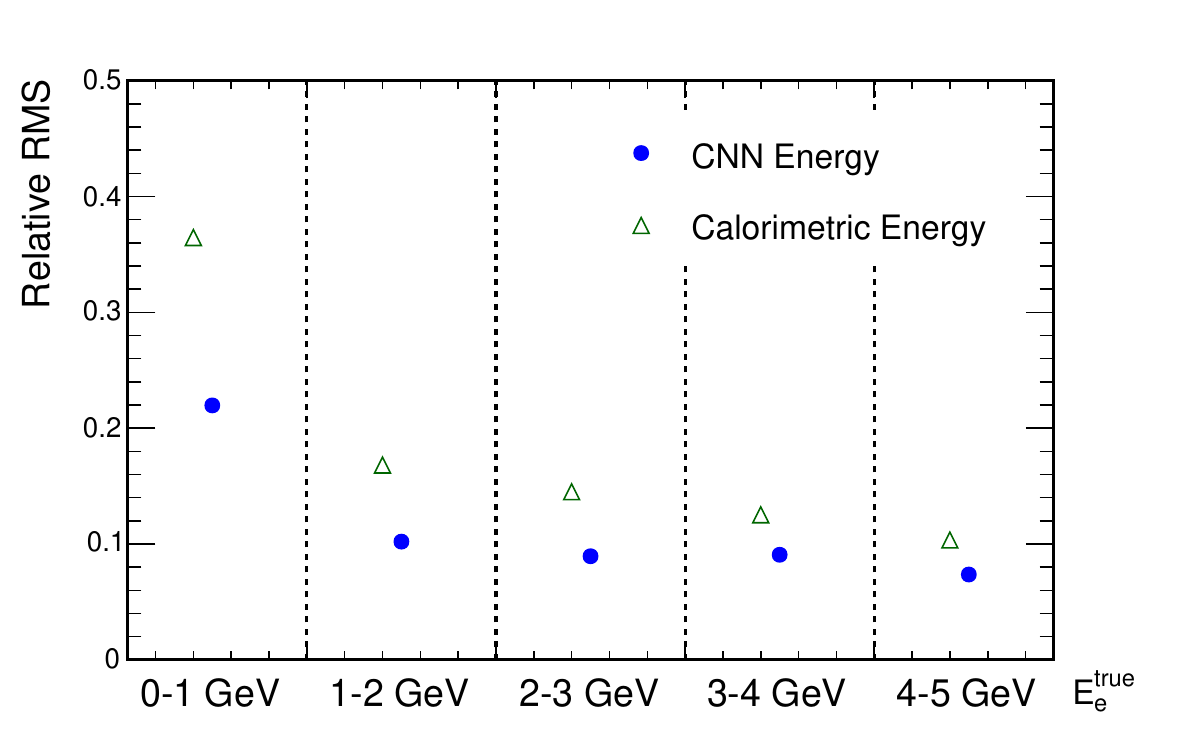}
    \caption{Means (left) and RMSs (right) of the Monte Carlo distributions of the ratios of reconstructed electron shower energy to true electron energy in $\nu_e$ CC events for different true electron energy bins ranging from 0 to 5 GeV. Shower energy is reconstructed by CNN (CNN Energy, blue) and summing the calibrated calorimetric energy in each cell (Calorimetric Energy, green). The neutrino oscillations are applied.}
    \label{fig:elecEResvsE}
\end{figure}

\begin{figure}[h]
    \centering
    \includegraphics[width=16cm]{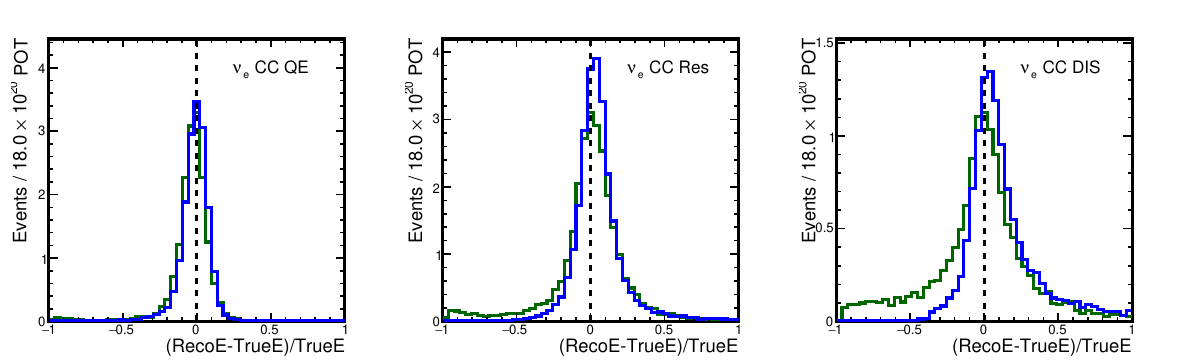}
    \caption{Monte Carlo distributions of ratios of reconstructed electron shower energy to true electron energy in $\nu_e$ CC events for QE, RES and DIS modes. Shower energy is reconstructed by CNN (CNN Energy, blue) and summing the calibrated calorimetric energy in each cell (Calorimetric Energy, green). The neutrino oscillations are applied.}
    \label{fig:elecEResMode}
\end{figure}

The reconstructed electron shower energy given by the regression CNN  (CNN Energy) is compared to the sum of the calibrated calorimetric energies (Calorimetric Energy) in electron showers.
In a simulated $\nu_e$ CC event, the most energetic shower matched to a true electron is chosen as the electron shower sample. The neutrino oscillation weights are applied to the $\nu_e$ CC events. True electron energy and reconstructed electron shower energy distributions in the FD are shown in Figure~\ref{fig:elecE}. The CNN electron shower energy is closer to the true electron energy than Calorimetric Energy. Overall $(E_{reco}-E_{true})/E_{true}$ for electron showers are shown in Figure~\ref{fig:elecERes}, with relative Gaussian resolutions of $8.2\%$ (CNN Energy) and $9.6\%$ (Calorimetric Energy) and relative RMS of $13.4\%$ (CNN Energy) and $15.2\%$ (Calorimetric Energy) .  Means and RMSs of $(E_{reco}-E_{true})/E_{true}$ in each 1-GeV-wide true electron energy bin, and histograms of $(E_{reco}-E_{true})/E_{true}$ in different interaction modes are shown in Figure~\ref{fig:elecEResvsE} and \ref{fig:elecEResMode}. One can find that the CNN Energy has better resolutions. The bias in CNN electron energy at low energies is caused by the small proportion of low energy electrons in the regular flux sample used for training.

\section{Summary}

We developed regression CNNs with direct pixel-level inputs for electron neutrino energy and electron shower energy reconstruction. This is an early effort and proof-of-concept of using CNNs to solve regression problems such as energy reconstruction, vertex reconstruction, and track parameter determination in HEP. It was found that the absolute scaled error provides a useful neural network loss function when the goal is to optimize energy resolution histograms as commonly used in HEP. We also describe the best training parameters found from hyperparameter search for this regression task. This work demonstrates that energy reconstruction tasks can be simplified without elaborate energy scale calibration and model-dependent fits. The performance of the CNN energy reconstruction was verified for different energy bins, different hadronic energy fractions and different interaction modes. In all cases, the regression CNN energy achieves superior performance compared with kinematics based energy reconstruction methods. The regression CNN also shows smaller systematic uncertainties from the simulation of neutrino interactions.

\section{Acknowledgments}

The authors thank the NOvA collaboration for use of its Monte Carlo simulation software and related tools. This work was supported by the US Department of Energy, the US National Science Foundation and University of California, Irvine. The DOE grant is ``FY 2018 Research Opportunities in High Energy Physics" (Comparative Review) Funding Opportunity Announcement [DE-FOA-0001781]. NOvA receives additional support from the Department of Science and Technology, India; the European Research Council; the MSMT CR, Czech Republic; the RAS, RMES, and RFBR, Russia; CNPq and FAPEG, Brazil; and the State and University of Minnesota. We are grateful for the contributions of the staff at the Ash River Laboratory, Argonne National Laboratory, and Fermilab. Fermilab is operated by Fermi Research Alliance, LLC under Contract No. De-AC02-07CH11359 with the US DOE.

\end{document}